\documentclass[runningheads,a4paper]{llncs}

\usepackage{amssymb}
\usepackage{graphicx}

\usepackage{url}
\urldef{\mailsa}\path|david.c.sterratt@ed.ac.uk, osorokin@inf.ed.ac.uk|
\urldef{\mailsb}\path|douglas.armstrong@ed.ac.uk|
\newcommand{\keywords}[1]{\par\addvspace\baselineskip
\noindent\keywordname\enspace\ignorespaces#1}

\def\citep{\cite}
\def\citet{\cite}

\usepackage[amssymb]{SIunits}
\addunit{\molar}{M}
\usepackage{mhchem}
\usepackage{amsmath}

\usepackage{color}

\newcommand{\Ichan}{\ensuremath{I^\mathrm{chan}}}
\newcommand{\Itchan}{\ensuremath{\tilde I^\mathrm{chan}}}
\newcommand{\Ipump}{\ensuremath{I^\mathrm{pump}}}

\newcommand{\labelgraphics}[2]{\parbox[t]{0.01\linewidth}{\vspace{0pt}\textsf{\textbf{#1}}}
  \parbox[t]{0.47\linewidth}{\vspace{0pt}#2}}

\begin{document}

\mainmatter  

\title{Integration of rule-based models and compartmental models of neurons}


%
%
\author{David C. Sterratt%
  \thanks{The research leading to these results has received funding
    from the European Union Seventh Framework Programme
    (FP7/2007-2013) under grant agreement nos. 241498 (EUROSPIN
    project), 242167 (SynSys-project) and 604102 (Human Brain
    Project).  We thank Anatoly Sorokin for his help with SpatialKappa
    and comments on an earlier version of the manuscript, and Vincent
    Danos for thought-provoking discussions.}
  \and Oksana Sorokina \and J. Douglas Armstrong}
%

\institute{School of Informatics, University of Edinburgh\\
10 Crichton St, Edinburgh EH8 9AB, UK\\
\mailsa\\
\mailsb
}

%
%

\toctitle{Lecture Notes in Computer Science}
\tocauthor{Authors' Instructions}
\maketitle

\begin{abstract}
  Synaptic plasticity depends on the interaction between electrical
  activity in neurons and the synaptic proteome, the collection of
  over 1000 proteins in the post-synaptic density (PSD) of synapses.
  To construct models of synaptic plasticity with realistic numbers of
  proteins, we aim to combine rule-based models of molecular
  interactions in the synaptic proteome with compartmental models of
  the electrical activity of neurons. Rule-based models allow
  interactions between the combinatorially large number of protein
  complexes in the postsynaptic proteome to be expressed
  straightforwardly. Simulations of rule-based models are stochastic
  and thus can deal with the small copy numbers of proteins and
  complexes in the PSD. Compartmental models of neurons are expressed
  as systems of coupled ordinary differential equations and solved
  deterministically.  We present an algorithm which incorporates
  stochastic rule-based models into deterministic compartmental models
  and demonstrate an implementation (``KappaNEURON'') of this hybrid
  system using the SpatialKappa and NEURON simulators.
\end{abstract}

\section{Introduction}
\label{neuron-kappa:sec:introduction}

The experimental phenomena of long term potentiation (LTP) and long
term depression (LTD) show that synapses can transduce patterns of
electrical activity on a timescale of milliseconds in the neurons they
connect into long-lasting changes in the expression levels of
neurotransmitter receptor proteins.  This synaptic plasticity plays a
crucial role in the development of a functional nervous system and in
encoding semantic memories (e.g.~motor patterns) and episodic memories
(experiences), converting stimuli lasting for seconds into memories
that last a lifetime \citep{MartEtal00syna}.

There are a number of computational models of how synaptic plasticity
arises from patterns of pre- and postsynaptic electrical activity, the
dynamics of $\alpha$-amino-3-hydroxy-5-methyl-4-isoxazolepropionic
acid receptors (AMPARs) and \textit{N}-methyl-D-aspartic acid
receptors (NMDARs), calcium influx through these receptors and
intracellular signalling in the postsynaptic density (PSD), a dense,
protein-rich structure attached to neurotransmitter receptors
\cite{BhalIyen99emer,SmolEtal06mode,LismZhab01mode,ZengHolm10effe}.
The level of detail of the molecular component of these models ranges
from deterministic simulations in one compartment
\citep{BhalIyen99emer} through stochastic models with coarse
granularity \citep{UrakEtal08requ} and, at the most detailed,
particle-based simulations in which the Brownian motion of individual
molecules is modelled \citep{StilBart01mont,ZengHolm10effe}.

The model with the greatest number of molecular species has 75
variables representing the concentrations of signalling molecules,
complexes of signalling molecules and phosphorylation states
\citep{BhalIyen99emer}.
This constitutes a small subset of the 1000 proteins identified in the
mouse postsynaptic proteome, the collection of proteins in the PSD
\citep{CollEtal06mole}. Even the subset of the postsynaptic proteome
containing proteins associated with membrane-bound neurotransmitter
receptors contains over 100 members \citep{PockEtal06prot}.  As these
proteins are particularly associated with synaptic plasticity, it
would be desirable to increase the number of proteins and complexes it
is possible to describe in simulations.

As each protein has a number of binding sites, a combinatorially large
number of complexes can arise, meaning that a correspondingly large
number of reactions are needed to describe the dynamics of all
possible complexes. By specifying rules whose elements are fragments
of complexes, rule-based languages and simulators
\citep{ChylEtal13inno}, such as Kappa \citep{DanoEtal07scal} or
BioNetGen \citep{FaedEtal09rule}, obviate the need to specify
reactions for all possible complexes.
These rules are simulated using a method similar to
Gillespie's stochastic simulation method for reactions
\citep{DanoEtal07scal}.
Kappa has been used to predict the sizes of clusters of proteins in
the postsynaptic proteome \citep{SoroEtal11towa}.

Compartmental models of electrical activity in neurons split the
neuronal morphology into a number (ranging from 1 to around 1000) of
compartments, and specify the dynamics of the membrane potential in
each compartment in terms of coupled ordinary differential equations
(ODEs) \citep{HodgHuxl52quant,SterEtal11prin}.  Quantities
beyond the membrane potential can also be modelled, e.g.\,
intracellular calcium concentration and concentrations of a few other
molecules such as buffers and pumps.
Various packages can generate and solve the equations underlying
compartmental models from various model description languages, for
example NEURON \citep{CarnHines06neur}, MOOSE \citep{SubhBhal08pymo}
and PSICS \citep{CannEtal10stoc}.

We present an algorithm which integrates rule-based models and
compartmental models of neurons.  To be sure of understanding a
simple, yet interesting, case, we limit ourselves to considering
isolated postsynaptic proteomes in a neuron of arbitrary
morphology. Although of interest, we do not consider diffusion of
molecules within the neuron. We implement the algorithm by
incorporating the SpatialKappa rule-based simulator \citep{SoroEtal13simul} into the
NEURON simulator \citep{CarnHines06neur}.  We validate the combined
simulator (``KappaNEURON'') against stand-alone NEURON, and demonstrate how the system
can be used to simulate complex models.

\section[Simulation method]{Simulation Method}
\label{sec-2}
\subsection{The System to Be Simulated}
\label{sec-2-1}

\begin{figure}[t]
  \centering
  \includegraphics[width=\linewidth]{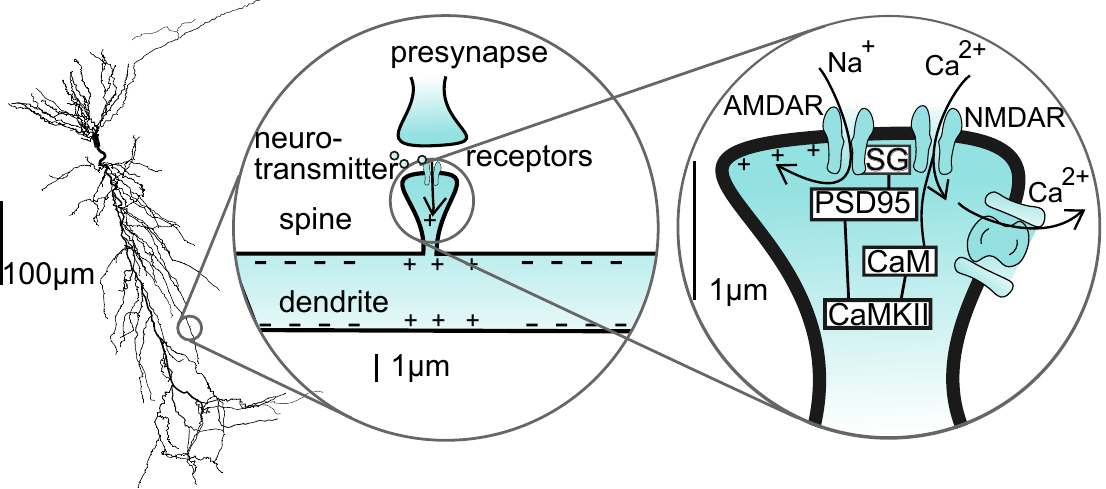}
  \caption{The system to be simulated depicted on the scale of a whole
    neuron (left), a spine (middle) and the postsynapse (right). Scale
    bars are approximate. In the postsynapse, the rectangular boxes
    represent intracellular molecules: SG -- stargazin; CaM --
    calmodulin; CaMKII -- Ca$^2+$/calmodulin-dependent protein kinase
    II.}
  \label{neuron-kappa:fig:kappa-in-neuron}
\end{figure}

An example of the type of system to be simulated is shown in
Fig.~\ref{neuron-kappa:fig:kappa-in-neuron}.  There is a hippocampal
CA1 pyramidal neuron (left) upon which are located a number of
synapses.  Excitatory synapses are generally located on synaptic
spines, small protuberances from the neuron whose narrow necks limit,
to an extent, diffusion of ions and molecules between the spine head
and the rest of the neuron (middle). The synapse contains a
postsynaptic proteome of arbitrary complexity (right). Firing in the
presynaptic neuron causes release of neurotransmitter from the
presynaptic bouton, which, after diffusing across the synaptic cleft,
binds to AMPARs and NMDARs. The AMPARs open and close on a
sub-millisecond timescale, allowing sodium ions to flow into the
cell. These ions charge the membrane locally and flow to other parts
of the neuron, where they also charge the membrane (middle,
Fig.~\ref{neuron-kappa:fig:kappa-in-neuron}).  The NMDARs open and
close with slower dynamics, and allow calcium ions to flow into the
spine. Inside the spine, the calcium ions bind to various proteins such as
calmodulin and the resulting calcium-calmodulin complex may then bind to
Ca$^2+$/calmodulin-dependent protein kinase II (CaMKII), initiating
signalling known to be critical for the induction of LTP and LTD.

We will first give the general set of equations that constitute a
deterministic description of this type of system, then apply the
general equations to a specific example, and next describe how the
equations are solved. Finally we will show how we incorporate
rule-based models and simulate the hybrid system. 

\subsection{Deterministic Description of Electrical and Chemical Activity in a Neuron}
\label{sec-2-2}

To describe electrical activity in the neuron, the neuron is split
into compartments, each of which should be small enough to be
approximately isopotential. Apart from the root compartment, which is
located in the soma, each compartment has a parent, and a compartment
may have one or more children. The equation for the membrane potential
$V_i$ in compartment $i$ derives from Kirchoff's current law:
\begin{equation}
C_i\frac{\mathrm{d}V_i}{\mathrm{d}t} = 
\sum_{j\in\mathcal{N}i} \frac{d_{ij}(V_j - V_i)}{4R_\mathrm{a}l_{ij}^2}
- \sum_S \left(
  \Ichan_{S,i} 
  + \Ipump_{S,i} 
\right)   
- \Ichan_{\mathrm{ns},i}\enspace .
\label{rb-compartmental-method:eq:3}
\end{equation}
The left hand side is the current per unit membrane area charging or
discharging the membrane; $C_i$ is the specific membrane capacitance
in compartment $i$.  The first term on the right hand side describes
current flow into compartment $i$ from its neighbours
$j\in\mathcal{N}_i$; $R_\mathrm{a}$ is the intracellular resistivity,
$l_{ij}$ is the path length between the midpoints of $i$ and each of
its neighbours $j$, and $d_{ij}$ is the mean diameter of the path. The
second term on the right hand side is the total transmembrane current
per unit area (referred to as current density) in compartment $i$
carried by various species of ion $S$ via ion channels
($\Ichan_{S,i}$) and membrane pumps ($\Ipump_{S,i}$), which act to
maintain concentration differences between the intracellular and
extracellular space. To represent ``non-specific'' ion currents whose
concentration is not accounted for, there is final term $\Ichan_{\mathrm{ns},i}$.  Here
the minus sign reflects the conventions that inward current is
negative and the extracellular space is regarded as electrical ground.

The current density carried by species $S$ through types $k$ of ion
channel is:
\begin{equation}
  \Ichan_{S,i}=\sum_kg_{ik}(O_{ik},t) f_{S,k}(V_i, [S]_i, [S]_\mathrm{o})
  \enspace ,
  \label{kappa-neuron-method:eq:2}
\end{equation}
where $g_{ik}$ is the conductance of ion channel type $k$ in
compartment $i$, which may be a function of time or a state variable
$O_{ik}$, and $f_{S,k}$ and is a function describing the $I$--$V$
characteristic of current flow of ions of type $S$ through channel
$k$, which may depend on $[S]_i$, the intracellular concentration of
$S$ in compartment $i$, and $[S]_\mathrm{o}$, the extracellular
concentration of $S$, which is assumed to be constant. A normalised
Goldmann-Hodgkin-Katz (GHK) current equation \citep{SterEtal11prin}
can be used for $f_{S,k}$. For channels through which calcium flows,
the typically large ratio between intracellular and extracellular
calcium concentrations means this function depends quite strongly on
the intracellular calcium concentration $[\mathrm{Ca}^{2+}]$, but in
channels not permeable to calcium it is usual to use a linear
approximation $V_i - E_k$, where $E_k$ is the reversal potential for
that channel. By removing the dependence on intracellular
concentrations $[S]_i$, this approximation allows currents carried by
ions other than calcium to lumped together in a nonspecific ion
category.

The state variable $O_{ik}$ is the number of type $k$ ion channels in
compartment $i$ which are in an open conformation. It is modelled as
the occupancy of a state of a Markov process with membrane
potential-dependent transition rates.  For small number of channels
the Markov process is simulated stochastically, but with large numbers
of channels the system is practically deterministic and the master
equation of the Markov process is simulated using
ODEs.

The dynamics of the intracellular concentrations of ions can be
modelled using further ODEs. The rate of change of $[S]_i$ depends on
$\Ichan_{S,i}$, the channel transmembrane current density carried by
$S$, and consumption and release by intracellular reactions:
\begin{equation}
  \frac{\mathrm{d}[S]_i}{\mathrm{d}t} = -\frac{a_i}{z_S Fv_i}\Ichan_{S,i} +
  \sum_r J_{S,r,i} \enspace ,
  \label{rb-compartmental-method:eq:1}
\end{equation}
where $a_i$ is the surface area of the compartment, $v_i$ is the
volume of the compartment, $z_S$ is the valency of ion $S$, and $F$ is
Faraday's constant.  The term $\sum_r J_{S,r,i}$ describes the net flux of $S$
due to intracellular reactions $r$. It arises from treating the
intracellular reactions in compartment $i$ as a set of kinetic
schemes:
\begin{equation}
  r: \ce{\textit{S} + \textit{T} <->[k_r][k_{-r}] \mathit{S\cdot{}T}} 
  \enspace .
\end{equation}
The flux of $S$ arising from this reaction would be:
\begin{equation}
  \label{rb-compartmental-method:eq:4}
  J_{S,r,i} = -k_r [S]_i[T]_i + k_{-r} [S\cdot T]_i \enspace .
\end{equation}
The pump current $\Ipump_{S,i}$ may be defined in terms of the flux of
a reaction $r$, for example:
\begin{equation}
  \label{rb-compartmental-method:eq:6}
  \Ipump_{S,i} = \frac{z_SFv_i}{a_i}J_{S,r,i} \enspace ,
\end{equation}
where the prefactor converts from flux to current. Thus the whole
electrical and molecular system is defined by a system of ODEs.

\subsection{Example Deterministic Description}
\label{rb-compartmental-method:sec:example-determ-descr}

To help understand the formalism above, we provide an example of a
simple one-compartment system; this will also be used as the
validation example in Section~\ref{neuron-kappa:sec:refer-simul}.
There is a single compartment whose membrane contains passive (leak)
channels, calcium channels, and a transmembrane calcium pump,
described by the kinetic scheme:
\begin{equation}
  \label{neuron-kappa:eq:1}
  \begin{aligned}
   \textrm{Ca binding:}\quad & \ce{P + Ca  ->[k_1] P.Ca} \\
   \textrm{Ca release:}\quad & \ce{P.Ca  ->[k_2] P}  \enspace ,
  \end{aligned}
\end{equation}
where Ca represents intracellular calcium, P represents a pump
molecule in the membrane, \ce{P.Ca} is the pump molecule bound by
calcium and $k_1$ and $k_2$ are rate coefficients. 

We substitute $z_\mathrm{Ca}=2$ and the flux of the ``Ca release''
reaction into Equation~(\ref{rb-compartmental-method:eq:6}) to obtain
the pump current:
\begin{equation}
  \label{rb-compartmental-method:eq:9}
  \Ipump_{\mathrm{Ca}} = k_2\frac{2Fv}{a}[\ce{P.Ca}] 
  = k_2\frac{2Fv}{a}([\ce{P}]_0 - [\ce{P}]) \enspace ,
\end{equation}
where we have used the fact that the total concentration of the pump
molecule $[\mathrm{P}]_0$ is the sum of the concentrations [P] and
[\ce{P.Ca}] of unbound and bound pump molecule. Since there is only a
single compartment, we have dropped subscripts.  The calcium channel
current $\Ichan_{\mathrm{Ca}}$ flowing into the compartment is
determined by Equation~(\ref{kappa-neuron-method:eq:2}) with a
constant conductance of magnitude $g_\mathrm{Ca}$, and the nonspecific
current is used for the passive channels so that
$\Ichan_{\mathrm{ns}}=g_\mathrm{pas}(V-E_\mathrm{pas})$, where
$E_\mathrm{pas}$ is the passive reversal potential.

To construct the ODEs corresponding the kinetic
scheme~(\ref{neuron-kappa:eq:1}) and the expression for the pump
current~(\ref{rb-compartmental-method:eq:9}),
Equation~(\ref{rb-compartmental-method:eq:4}) is applied to the scheme
to give fluxes, which, along with $\Ichan_\mathrm{Ca}$, 
$\Ipump_\mathrm{Ca}$ and $\Ichan_{\mathrm{ns}}$, are substituted in
Equations~(\ref{rb-compartmental-method:eq:3})
and~(\ref{rb-compartmental-method:eq:1}) to give:
\begin{align}
  C\frac{\mathrm{d}V}{\mathrm{d}t} & 
  =  -g_\mathrm{Ca}f_\mathrm{Ca}(V,[\ce{Ca}], [\ce{Ca}]_\mathrm{o}) - k_2\frac{2Fv}{a}([\ce{P}]_0 -
  [\ce{P}]) - g_\mathrm{pas}(V-E_\mathrm{pas})   
  \label{rb-compartmental-method:eq:5}
 \\
  \frac{\mathrm{d}[\ce{Ca}]}{\mathrm{d}t} & 
  = -\frac{a}{2Fv}g_\mathrm{Ca}f(V,[\ce{Ca}], [\ce{Ca}]_\mathrm{o}) 
  - k_1[\ce{Ca}][\ce{P}]  
  \label{rb-compartmental-method:eq:7}
  \\
  \frac{\mathrm{d}[\ce{P}]}{\mathrm{d}t} & 
  = -k_1[\ce{Ca}][\ce{P}]+ k_2([\ce{P}]_0 - [\ce{P}])
  \label{rb-compartmental-method:eq:8}
  \enspace .
\end{align}

The notional volume $v$ may describe the volume of a thin submembrane
shell rather than the volume of the whole compartment.  We assume that
$v$ is the volume of the whole cylindrical compartment so that
$a/v=4/d$, where $d$ is the diameter of the compartment.

\subsection{Simulation of Deterministic Variables}

Simulators of deterministic electrical and chemical activity in
neurons, such as NEURON \citep{CarnHines06neur}, solve the coupled
ODEs by gathering the variables $V_i$, $[S]_i$ and other state
variables into one state vector $\vec{x}$ and solving the ODE system:
\begin{equation}
  \label{neuron-kappa:eq:5}
  \frac{\mathrm{d} \vec{x}}{\mathrm{d} t} = \vec{G}(\vec{x}) + \vec{b}(t)
  \enspace ,
\end{equation}
where $\vec{G}(\vec{x})$ is the rate of change of each state variable
and $\vec{b}(t)$ is a time dependent forcing input.  In principle
$\vec{G}(\vec{x})$ depends on all variables, though the structure of
compartmental models means each element of $\vec{G}(\vec{x})$ depends
on only a few elements of $\vec{x}$.  These equations can be solved by
implicit Euler integration, which, although not providing the
second-order accuracy of more advanced schemes, does give guarantees
about numerical stability and is used by default in NEURON
\citep{McDoEtal13reac}.  In implicit Euler the derivative is evaluated
at $t+\Delta t$, the end of the time step:
\begin{equation}
  \frac{\vec{x}(t+\Delta t) - \vec{x}(t)}{\Delta t} =
  \vec{G}(\vec{x}(t+\Delta t))  + \vec{b}(t+\Delta t) \enspace .
  \label{neuron-kappa:eq:6}
\end{equation}
Taylor expanding the right hand side of~(\ref{neuron-kappa:eq:6}) in
$\Delta t$ and rearranging gives the equation for one update step:
\begin{equation}
  \label{neuron-kappa:eq:7}
  \vec{x}(t+\Delta t) = \vec{x}(t) 
  + \left(I - \frac{\partial \vec{G}}{\partial \vec{x}}\Delta
    t\right)^{-1}
  \left(\vec{G}(\vec{x}(t)) + \vec{b}(t)\right)\Delta t \enspace ,
\end{equation}
where $\partial \vec{G}/\partial \vec{x}$ is the Jacobian matrix at
time $t$, which can be computed numerically, or analytically for
efficiency. To optimise simulation speed, the time step $\Delta t$ may
vary depending on the rate of change of the variables; but whatever
the value of $\Delta t$, all variables are updated simultaneously.

\subsection{Modifications to Accommodate Rule-Based Simulation}
\label{rb-compartmental-method:sec:modif-accomm-rule}

To combine fixed-step simulation of continuous variables described by
ODEs with discrete variables described by stochastic, rule-based
models, we use principles akin to those used in hybrid simulations of
systems of chemical reactions \citep{KiehEtal04hybr}. The state
variables (elements of $\vec{x}$) are partitioned into continuous
variables, which are updated at fixed intervals of $\Delta t$ by an
ODE solver, and discrete variables, which are updated asynchronously
by the rule-based solver, as outlined in
Appendix~\ref{neuron-kappa:sec:kappa-integr-meth}. Some ``bridge''
variables are referred to by both solvers. The combined simulation
algorithm must ensure that the two solvers are synchronised
appropriately and that conversions between continuous and discrete
quantities are made.

For simulations combining molecular and electrical activity
(e.g.~Fig.~\ref{neuron-kappa:fig:kappa-in-neuron}) the membrane
potential would be a continuous variable, intracellular molecules such
as calmodulin and CaMKII would be stochastic variables, and the
intracellular calcium in the spine would be a stochastic bridge
variable.

\paragraph{Conversions} In deterministic simulations of biochemical
reactions in neurons (e.g.~\cite{BhalIyen99emer}) a molecular species
or ion $S$ is represented by an intensive quantity -- its concentration
$[S]$; whereas in stochastic simulations it is represented by an
extensive quantity -- the number of molecules $|S|$ in the volume $v$ in
which $S$ exists. Thus to compare the deterministic (ODE) and
stochastic (rule-based) parts of the simulation, intensive and
extensive quantities need to be interconverted using Avogadro's
constant $N_\mathrm{A}$:
\begin{equation}
  \label{neuron-kappa:eq:8}
  |S| = N_\mathrm{A}v[\ce{S}] \enspace .
\end{equation}
Rate coefficients for reactions based on concentrations must also be
converted to ones appropriate for species number for use in the
rule-based simulator's rules.  To derive the conversion formula,
consider a bimolecular kinetic scheme in which $k$ is the rate
coefficient, i.e.~\ce{S + T ->[k] S.T}. Typical units for $k$ are
\reciprocal\molar\reciprocal\second. The kinetic scheme can be
expressed as an ODE
$\mathrm{d}[\ce{\mathit{S}.\mathit{T}}]/\mathrm{d}t = k[S][T]$.
Converting the concentrations according to~(\ref{neuron-kappa:eq:8})
yields an equivalent ODE whose variables are numbers of molecules:
$\mathrm{d}|\ce{\mathit{S}.\mathit{T}}|/\mathrm{d}t = \gamma|S||T|$,
where $\gamma = k/N_\mathrm{A}v$ is the converted rate coefficient and
has units \reciprocal\second. In general for an equation with $n$ reactants, the
relation between rate coefficients for numbers of molecules and
concentrations is $\gamma = (N_\mathrm{A}v)^{-n+1}k$.  State variables
(for example the state of channels) may also be controlled by the
rule-based simulator, and here other conversion formulae apply.

\paragraph{Creation and Destruction Rules for Bridging Variables} To
simulate the channel currents in the rule-based solver we need to
write creation or destruction rules that are equivalent to
$\Ichan_{S,i}$ in Equation~(\ref{rb-compartmental-method:eq:1}). These
rules are:
\begin{align}
  \label{rb-compartmental-method:eq:2}
  & \mbox{ } \ce{->[-a_i\Itchan_{S,i}N_\mathrm{A}/z_SF] \mathit{S}} &
  \mbox{if } \Itchan_{S,i} < 0 \\
  & \ce{\mathit{S} ->[a_i\Itchan_{S,i}N_\mathrm{A}/z_SF]}  &
  \mbox{if } \Itchan_{S,i} > 0.
\end{align}
Here $\Itchan_{S,i}$ can be an expression that references continuous or
discrete variables.

\begin{figure}[t]
  \centering
  \includegraphics{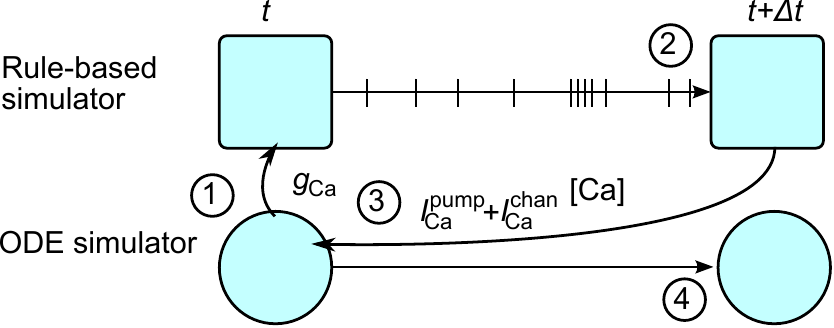}
  \caption{The update and synchronisation method. See text for
    explanation.}
  \label{rb-compartmental-method:fig:integration}
\end{figure}

\paragraph{Update and Synchronisation} The procedure for updating the
time from $t$ to $t+\Delta t$
(Fig.~\ref{rb-compartmental-method:fig:integration}) is:
\begin{enumerate}
\item Pass all relevant continuous variables, e.g. conductances and
  voltages needed to compute $\Ichan_{S,i}$ in the rule-based
  simulator.
\item Run the rule-based simulator from $t$ to $t + \Delta t$.
\item Compute the net change $\Delta S_i^\mathrm{tot}$ in the total
  number of each bridging species $S$ (including in any complexes) in
  compartment $i$ over the time step and convert each change back to a
  current: $\Ichan_{S,i} + \Ipump_{S,i} =-\Delta S^\mathrm{tot}_i
  z_SF/a_iN_\mathrm{A}$.  For each membrane potential $V_i$, set the
  corresponding element of $\vec{b}(t)$ equal to
  $-(1/C_i)\sum_S({\Ichan_{S,i} + \Ipump_{S,i}})$ (see second term on
  right of Equation~\ref{rb-compartmental-method:eq:3}).
\item Update the continuous variables according to the update
  step~(\ref{neuron-kappa:eq:7}).
\end{enumerate}
When running the rule-based model, it will not stop precisely on the
boundary of the time step since the update times are generated
stochastically. To deal with this problem, the time of the next event
in the rule-based component is computed before updating the
variables. As soon as the next event time is after the end of the
deterministic step, that event is thrown away, as justified in
Appendix~\ref{neuron-kappa:sec:details-kappa-update}.

\begin{figure}
\begin{verbatim}
## File caPump.ka - Simple calcium pump

## Agent declarations, showing the agent names and binding sites
%agent: ca(x)   # Calcium with binding site
%agent: P(x)    # Pump molecule with binding site

## Variable declarations
%var: 'vol' 1          # Volume in um3 
%var: 'NA'  6.02205E23 # Avagadro's constant
# Concentration of one agent in the volume in mM 
%var: 'agconc' 1E18/('NA' * 'vol') 
# Rate constants in /mM-ms or /ms, depending on the number of
# complexes on LHS of rule
%var: 'k1' 0.001       # /mM-ms 
%var: 'k2' 1           # /ms

## Rules 
# Note the scaling of the rate constant of the bimolecular reaction
'ca binding' ca(x), P(x)     -> ca(x!1), P(x!1) @ 'k1' * 'agconc'
'ca release' ca(x!1), P(x!1) -> P(x)            @ 'k2'

## Initialisation of agent numbers
# Overwritten by NEURON but needed for SpatialKappa parser
%init:  1000 ca(x) 
%init: 10000 P(x)

## Observations
%obs: 'ca'   ca(x)           # Free Ca
%obs: 'P-Ca' ca(x!1), P(x!1) # Bound Ca-P
%obs: 'P'    P(x)            # Free P
\end{verbatim}
  \caption{Example of a Kappa file for a simple calcium
    pump~(\ref{neuron-kappa:eq:1}) simulated in a volume of
    1\micro\meter\cubed. Note the conversion of the forward rate
    coefficient from units of
    \milli\reciprocal\molar\usk\milli\reciprocal\second\ to
    \milli\reciprocal\second. For an introduction to the Kappa
    language, see the short description at
    \protect\url{http://www.kappalanguage.org/syntax.html}. }
\label{neuron-kappa:fig:kappa-pump}
\end{figure}

\begin{figure}
  \centering
\begin{verbatim}
from neuron import *
import KappaNEURON

## Create a compartment
sh = h.Section()
sh.insert('pas')                # Passive channel
sh.insert('capulse')            # Code to give Ca pulse
# This setting of parameters gives a calcium influx and pump
# activation that is scale-independent
sh.gcalbar_capulse = gcalbar*sh.diam

## Define region where the dynamics will occur ('i' means intracellular) 
r = rxd.Region([sh], nrn_region='i')

## Define the species, the ca ion (already built-in to NEURON), and the
## pump molecule. These names must correspond to the agent names in
## the Kappa file.
ca = rxd.Species(r, name='ca', charge=2, initial=0.0)
P  = rxd.Species(r, name='P',  charge=0, initial=0.2)

## Create the link between the Kappa model and the species just defined 
kappa = KappaNEURON.Kappa(membrane_species=[ca], species=[P], 
                          kappa_file='caPump.ka', regions=[r])

## Transfer variable settings to the kappa model
vol = sh.L*numpy.pi*(sh.diam/2)**2
kappa.setVariable('k1',  47.3)
kappa.setVariable('k2',  gamma2)
kappa.setVariable('vol', vol)

## Run
init()
run(30)
\end{verbatim}
  \caption{Extract of Python code to link the Kappa file shown in
    Fig.~\ref{neuron-kappa:fig:kappa-pump} into a compartment in
    NEURON.}
  \label{neuron-kappa:fig:kappa-neuron}
\end{figure}

\section[Implementation]{Implementation}
\label{neuron-kappa:sec:implementation}

We have implemented the algorithm described in the previous section by
linking the Java-based SpatialKappa implementation of the Kappa
language \citep{SoroEtal13simul} to version 7.4 of NEURON, which
allows reaction-diffusion equations to be specified in Python
\citep{McDoEtal13reac}.  Our implementation (``KappaNEURON'') is
available at
\url{http://github.com/davidcsterratt/KappaNEURON}.
We have used the
py4j\footnote{\url{http://py4j.sourceforge.net/}}
package to extend the SpatialKappa simulator so its Java objects can
be accessed in Python. The wrapper system in NEURON~7.4 allows us to
override NEURON's built-in fixed solve callback function with one that
calls the SpatialKappa simulator at each time step, as described in
the previous section.

In order to specify a model the Kappa component is specified in a
separate file.  Fig.~\ref{neuron-kappa:fig:kappa-pump} shows an
example of a simple calcium pump specified in Kappa. This file is then
linked into the NEURON simulation as demonstrated in the Python code
in Fig.~\ref{neuron-kappa:fig:kappa-neuron}. The mechanisms specified
in the Kappa file take over all of NEURON's handling of molecules in
the cytoplasm of chosen sections.

\section{Results}
\label{kappa-neuron:sec:results}

\subsection{Validation}
\label{neuron-kappa:sec:refer-simul}

\begin{figure}[t]
  \labelgraphics{A}{\includegraphics{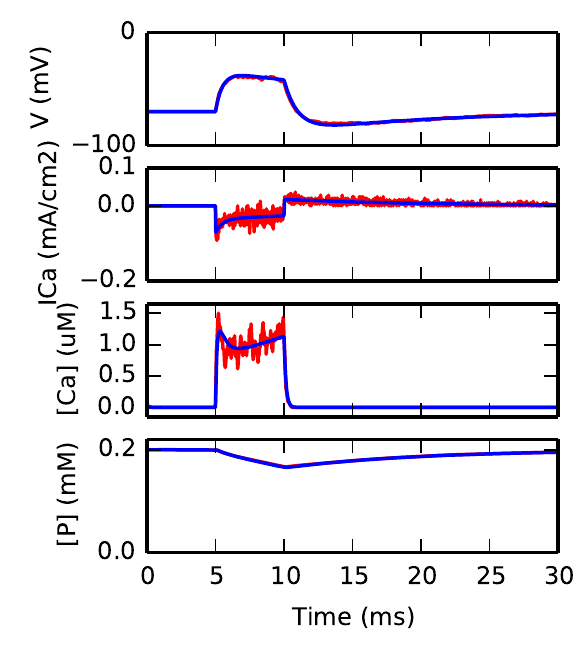}}
  \labelgraphics{B}{\includegraphics{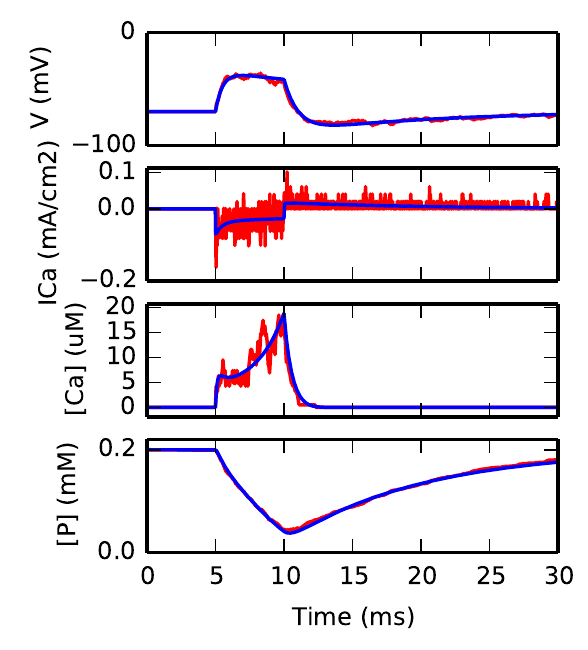}}
  \caption{Reference simulations. \textbf{A,}~Traces generated
    by NEURON with NMODL (blue) and KappaNEURON (red) when the
    diameter is 1\micro\meter. \textbf{B,}~The same simulations but
    with a diameter of 0.2\micro\meter. There is more noise evident in
    the combined simulation due to the smaller number of ions
    involved.}
\label{neuron-kappa:fig:refsim1}
\end{figure}

We validated our implementation by comparing the results of simulating
ODE and rule-based versions of the model described in
Section~\ref{rb-compartmental-method:sec:example-determ-descr} using
standard NEURON and KappaNEURON
respectively. Fig.~\ref{neuron-kappa:fig:refsim1}A shows the
deterministic solution of the system of
ODEs~(\ref{rb-compartmental-method:eq:5})--(\ref{rb-compartmental-method:eq:8})
(blue) and a sample rule-based solution using the Kappa rules in
Fig.~\ref{neuron-kappa:fig:kappa-pump} (red) from a single compartment
with diameter 1\micro\meter\ and length 1\micro\meter, giving a volume
within the range 0.01--1\micro\meter\cubed{}
typical of spine heads in the vertebrate central nervous system
\citep{HarrKate94dend}.
The calcium conductance $g_\mathrm{Ca}$ is zero apart from during a
pulse lasting from 5--10\milli\second\ when an inward calcium current
begins to flow ($\Ichan_{\mathrm{Ca}}$ is negative).  This causes a
sharp rise in intracellular calcium concentration [Ca], which, because
of the GHK current equation, reduces the calcium current slightly,
accounting for the initially larger magnitude of the calcium
current. As the calcium concentration increases, it starts binding to
the pump molecules, depleting the amount of the free pump molecules
[P]. Once the calcium channels close ($g_\mathrm{Ca}=0$), the calcium
influx stops, and the remaining free calcium is taken up quickly by
the pumps. The pump-calcium complex dissociates at a slower rate,
leading to a positive (outwards) calcium current.
The stochastic traces (red) are very similar to their deterministic
counterparts (blue) apart from some random fluctuations, particularly
in the trace of calcium. This agreement, along with a suite of simpler
tests included with the source code, indicates that the implementation
is correct.  

Fig.~\ref{neuron-kappa:fig:refsim1}B shows deterministic (blue) and
stochastic (red) simulations in a spine with a diameter of
0.2\micro\meter\ (i.e.~1/25 of the volume of the simulation in
Fig.~\ref{neuron-kappa:fig:refsim1}A). The shape of the traces differs
due to the change in surface area to volume ratio. Due to the smaller
volume and hence smaller numbers of ions involved, the fluctuations
are relatively bigger.

\subsection{Demonstration Simulation}
\label{neuron-kappa:sec:demonstrations}

To demonstrate the utility of integrated electrical and rule-based
neuronal models, we constructed a model of a subset of the synaptic
proteome, with a focus on the signal processing at the early stages of
the CaM-CaMKII pathway. We encoded in Kappa published models of: the
dynamics of NMDARs \citet{UrakEtal08requ}; binding of calcium with
calmodulin and binding of calmodulin-calcium complexes to CaMKII
\citet{PepkEtal10dyna}; and binding of calcium to a calbindin buffer
\citet{FaasEtal11calm}. We embedded these linked models into a simple
model of a synaptic spine, comprising head and neck compartments,
connected to a dendrite. As well as the NMDARs, modelled in Kappa,
there were AMPARs in the spine head, and backpropagating spikes were
modelled by inserting standard Hodgkin-Huxley ion channels
\citep{HodgHuxl52quant} in the dendritic membrane.  To emulate
spike-timing dependent synaptic plasticity protocols
\citep{MarkEtal97regul}, a train of 10 excitatory postsynaptic
potentials (EPSPs) were induced in the synapse in the spine head at
20Hz, each of which was followed by an action potential. There were
also 50 other synaptic inputs onto the dendrite, though these did not
contain the rule-based model.
\begin{figure}[t]
  \centering
  \labelgraphics{A}{\includegraphics{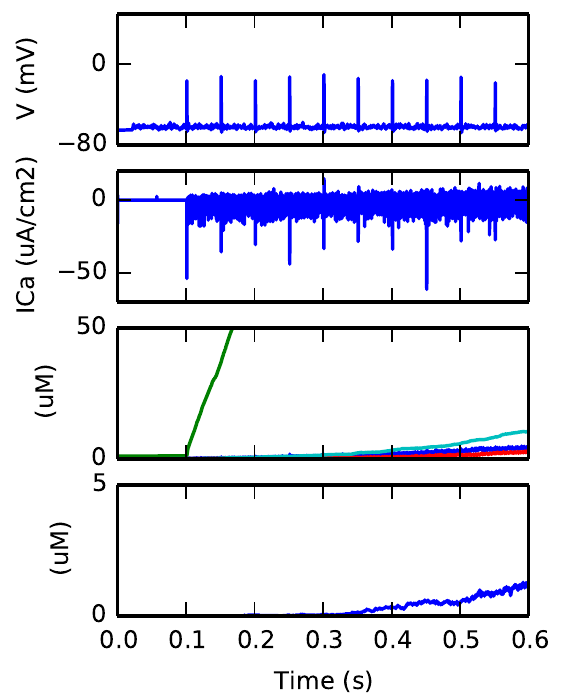}}
  \labelgraphics{B}{\includegraphics{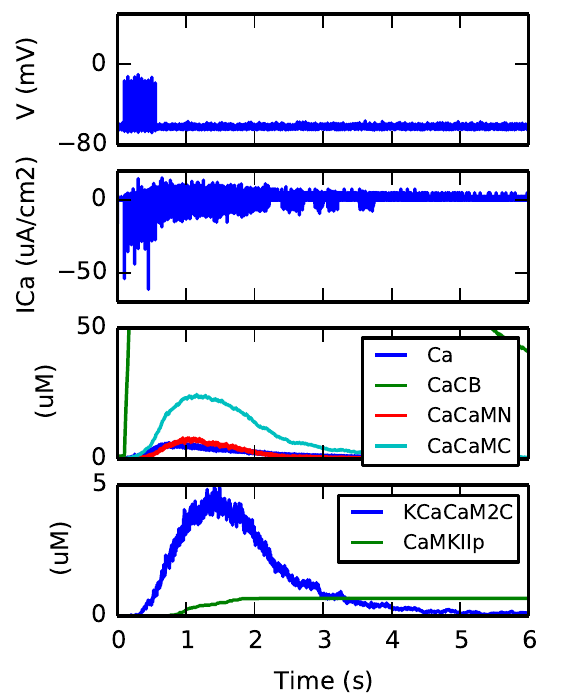}}
  \caption{Demonstration simulations. \textbf{A,}~The first 600ms of
    the simulation. \textbf{B,}~The first 6000ms of the
    simulation. Labels: Ca, free calcium; CaCB, calcium bound to
    calbindin; CaCaMN, calcium bound to the N lobe of calmodulin;
    CaCaMC, calcium bound to the C lobe of calmodulin; KCaCaM2C,
    CaMKII bound to calmodulin with two calcium ions on its C lobe;
    and CaMKIIp, phosphorylated CaMKII.}
  \label{neuron-kappa:fig:demo-sims}
\end{figure}

Fig.~\ref{neuron-kappa:fig:demo-sims} shows results from one
simulation at short (0.6s) and long (6s) time scales.  The first
stimulation of the detailed synapse paired with action potential
initiation occurs at 0.1s, as can be seen by the voltage trace. The
stimulation releases glutamate, which binds to AMPARs (which remain
open for a few milliseconds) and NMDARs (which remain open for 100s of
milliseconds).  Due to the backpropagating action potentials releasing
the voltage-dependent block of NMDARs, there are peaks in the calcium
current ($I_\mathrm{Ca}$) at the same time as the action
potentials. The calcium entering through the NMDARs binds to the
calbindin and calmodulin (CaM) buffers. The CaM-Ca$^{2+}$ complex can
bind to CaMKII, the rate depending on which of the C and N lobes of
CaM the Ca$^{2+}$ are bound to. This CaMKII-CaM-Ca$^{2+}$ complex can
then be phosphorylated, leading to a long lasting elevation in its
level. This will then to phosphorylate stargazin, which will help to
anchor AMPARs in the membrane, thus contributing to LTP.

\section{Discussion}
\label{neuron-kappa:sec:discussion}

We have presented a method for integrating a stochastically simulated
rule-based model of proteins in a micron-sized region of a neuron into
a compartmental model of electrical activity in the whole neuron.  The
rule-based component allows the biochemical interactions between
binding sites on proteins to be specified using a tractable number of
rules, with the simulator taking on the work of tracking which
complexes are present at any point during the simulation.

Our approach is similar to that of Kiehl et al.~\citep{KiehEtal04hybr}
who simulated chemical reactions with a hybrid scheme.  However their
integration scheme synchronised at every discrete event, whereas in
ours, synchronisation is driven only by the time step of the
continuous simulator. This principle is appropriate for neural
systems, in which we can expect many discrete events per time step.

Recently Mattioni and Le Nov\`ere \citet{MattNove13inte} have
integrated the ECell simulator with NEURON. Our approach is similar to
theirs, though with two differences. Firstly we have integrated a
rule-based simulator. This has the advantage that the interactions
between the combinatorially large numbers of complexes present in the
PSD can be specified using a tractable number of rules, though this
limits the simulation method to the sequential style of Gillespie's
stochastic simulation algorithm, and does not allow for any of the
approximations that increase the algorithm's efficiency. Secondly
Mattioni and Le Nov\`ere get the ODE-based solver to handle calcium,
whereas we handle it in the rule-based solver. Our approach is less
efficient computationally, but it ensures that all biochemical
quantities are consistent and avoids having to make any assumptions
about the relative speeds of processes.

Our approach allows us to model at a considerable level of detail. For
example the conformation of NMDARs may be part of the biochemical
model, allowing proteins in the PSD (e.g.~calmodulin bound to calcium)
to modulate the state of the channel \citep{UrakEtal08requ}. We can
also use one rule-based scheme to model both the presynapse and the
postsynapse, which could help to understand transynaptic signalling
via molecules such as endocannabinoids \citep{CastEtal12endo}.

Our implementation of our algorithm is publicly available
(KappaNEURON;
\url{http://github.com/davidcsterratt/KappaNEURON})
and under development.  The next major feature planned is 
making available to NEURON SpatialKappa's capability of simulating
rule-based models with voxel-based diffusion.

\appendix
\section{Appendix}
\label{neuron-kappa:sec:method}

\subsection{Kappa Simulation Method}
\label{neuron-kappa:sec:kappa-integr-meth}

To understand the asynchronous nature of the Kappa simulation method,
we first illustrate Gillespie's direct method \citep{Gill77exac} by
applying it to the kinetic scheme description of a calcium pump shown
in Fig.~\ref{neuron-kappa:fig:kappa-pump}A.  Here Ca represents
intracellular calcium, P represents a pump molecule in the membrane,
\ce{P.Ca} is the pump molecule bound by calcium and $k_1$ and $k_2$
are rate coefficients, which are rescaled to the variables $\gamma_1$
and $\gamma_2$ as explained in
Section~\ref{rb-compartmental-method:sec:modif-accomm-rule}. To apply
the Gillespie method to this scheme:
\begin{enumerate}
\item Compute the \emph{propensities} of the reactions $a_1 = \gamma_1
  |\ce{Ca}||\ce{P}|$ and $a_2 = \gamma_2 |\ce{Ca.P}|$
\item The total propensity is $A = a_1+a_2$
\item Pick reaction $\mathrm{R}_i$ with probability $a_i/A$
\item Pick time to reaction $T = -(\ln r)/A$, where $r$ is a random
  number drawn uniformly from the interval $(0,1)$.
\item Goto 1
\end{enumerate}
Kappa uses an analogous method, but applied to rules that are
currently active. Both methods are event-based rather than time-step
based.

\subsection{Justification for Throwing Away Events}
\label{neuron-kappa:sec:details-kappa-update}

To justify throwing away events occurring after a time step ending at
$t+\Delta t$, we need to show that the distribution of event times
(measured from $t$) is the same in two cases:
\begin{enumerate}
\item The event time $T$ is drawn from an exponential distribution
  $A\exp(-AT)$ (for $T>0$), where $A$ is the propensity.
\item An event time $T_0$ is drawn as above. If $T_0<\Delta t$, accept
  $T=T_0$ as the event time. If $T_0\geq\Delta t$, throw away this
  event time and sample a new interval $T_1$ from an exponential
  distribution with a time constant of $A$, i.e.\ $A\exp(-AT_1)$. Set
  the event time to $T=\Delta t + T_1$.
\end{enumerate}
In the second case, the overall distribution is:
\begin{equation}
  \begin{aligned}
  P(\mbox{event at } T < \Delta t) & = A\exp(-AT) \\
  P(\mbox{event at } T \geq \Delta t) &= P(\mbox{survival to } \Delta t)
  P(\mbox{event at } T_1) \\
  & =  \exp(-A\Delta t) A\exp(-A(T-\Delta t))  \\
  & = A\exp(-AT) 
  \end{aligned}
\end{equation}
Here we have used $T_1 = T - \Delta t$. Thus the distributions are the
same in both cases.

\bibliography{rb-compartmental-method}

\end{document}